\documentclass[twocolumn,aps]{revtex4}

\usepackage{amsfonts}
\usepackage{amsmath}
\usepackage{bm}
\usepackage{graphicx}
\usepackage{xcolor}
\usepackage{color}
\usepackage{amssymb}
\usepackage{natbib}
\usepackage{mwe}
\usepackage{caption}
\usepackage{subcaption}
\usepackage{indentfirst}
\usepackage{booktabs}
\usepackage{listings}
\usepackage{tabularx,array}
\usepackage{multirow}
\usepackage{hyperref}

\def \beq {\begin{equation}}
\def \eeq {\end{equation}}
\def \bea {\begin{eqnarray}}
\def \eea {\end{eqnarray}}
\def \bfig {\begin{figure}}
\def \efig {\end{figure}}

\DeclareMathAlphabet{\mathpzc}{OT1}{pzc}{m}{it}

\hypersetup{
     pdfpagemode = {UseOutlines},
     bookmarksopen,
     pdfstartview = {FitH},
     colorlinks,
     linkcolor = {black},
     citecolor = {black},
     urlcolor = {black}
}

\begin{document}

\title{Simulations Campaign of the Turbulent Diffusion at Tokamaks Edges}

\author{L. Scarivaglione, F. Valentini, and S. Servidio}

\affiliation{Dipartimento di Fisica, Universit\`a della Calabria, I-87036 Cosenza, Italy}

\begin{abstract}
The understanding of cross-field transport is crucial in order to optimize the properties of magnetic confinement in modern fusion devices. In this work, a two-dimensional, simplified model is used to study the turbulent dynamics in the region of the scrape-off layer. We show how the numerical model, based on the reduced Braginskii equations, is able to describe the formation and the evolution over time of blobs structures. We study these complex dynamics by using both classical Eulerian analysis and the Lagrangian approach, by varying the ambient conditions of the plasma. We have found that both the magnetic shear and the plasma profiles are crucial for the properties of the transport. By following Lagrangian tracers, we observed diffusive transients for the radial transport, at lengthscales larger than the typical blob size. This work is relevant for the comprehension of the turbulent transport at the edges of tokamak devices.

\end{abstract}

\date{\today}

\maketitle

\section{Introduction}
Experimental observations reveal that in plasma fusion devices the region near the external walls is characterized by the presence of high-density (high-temperature) structures, called ``filaments'' or ``blobs'' \cite{BoedoEA14, SechrestEA11, GarciaEA07, DevynckEA06, KallmanEA10}. These structures propagate outward from the inner region of the torus to the region between the plasma edge and the wall of the device, leading to particles and heat loss \cite{Endler1999, AntarEA01, MilitelloEA16}. This outer region is known as the scrape-off layer (SOL), characterized by open magnetic field lines \cite{Bates1996}. Understanding the particle transport mechanism in the SOL is of extreme importance in the perspective of achieving optimal conditions for the production of energy from nuclear power plants.

The blobs in the SOL appear as mushrooms with an elongated tail, like a kind of small discharge -- a ``burst'' \cite{Garcia_2009}. Several studies reveal that the basic mechanism responsible for the radial transport of blobs is due to a gradient of the magnetic field \cite{Krasheninnikov_2001, D'IppolitoEA02}, whose strength is inversely proportional to the major radius of the torus. This causes (1) drift in the curvature of the particles which induces a charge imbalance, (2) a subsequent polarization in the poloidal direction, and (3) radial advection towards the walls. In this scenario, particles are advected across the magnetic field lines due to the electric drift velocity $\mathbf{u}_E = \frac{\mathbf{E} \times \mathbf{B}}{B^2}$, leading to a radial loss of matter with a subsequent waist of the confinement properties \cite{Lawson_1957, Garbet_2006}.

The above bursty blobs are the outcome of instabilities in regions of open magnetic field lines, where the magnetic drift is not fully stabilized. However, the turbulent transport can be greatly reduced in the presence of a stable shear flow \cite{Terry_2000}. The sheared flow is due to the radial variation of the poloidal component of the electric cross-drift. When the shearing rate is comparable to the nonlinear turbulence decorrelation rate, the turbulent motion is suppressed \cite{BiglariEA1990, RitzEA1990, VanOostEA03, WareEA1998, HidalgoEA03, Horacek_2006, Fundamensky_2008}. There is therefore an interchange of kinetic energy between the fluctuating motion and the sheared poloidal flow \cite{Diamond_1991}, where the transport exhibits a bursty behavior, alternating quiet and violent-transport times \cite{BianEA03, ServidioEA08}.

The understanding of cross-field transport is crucial in order to optimize the properties of magnetic confinement in modern fusion devices, and numerical models are crucial for the comprehension of radial transport. In this work, a two-dimensional, simplified model is used to study the turbulent dynamics in the region of the scrape-off layer. We show how the numerical model, based on the reduced Braginskii equations, is able to describe the formation and the evolution over time of blobs structures. We study these complex dynamics by using both classical Eulerian analysis and the Lagrangian approach, by varying the ambient conditions of the plasma. 

The paper is structured as follows. In Section \ref{sec:model} we briefly present the governing equations of our model. Section \ref{sec:numerical} focuses on the description of the numerical code that we developed to solve the equations. We present the simulation campaign, varying important parameters that allow us to obtain different turbulent regimes. In Section \ref{sec:results} we present the simulation results, from both the Eulerian and the Lagrangian perspectives. Discussions and conclusions are provided in the last section.


\section{The model}
\label{sec:model}
Since at the edge of tokamak devices, the plasma is relatively cold, and the densities are still very high, the huge collisionality allows the use of a reduced fluid description \cite{Chen_1984, MosettoEA13}. The plasma can therefore be described by considering a few moments of the particle distribution function, leading to a set of fluid equations, such as the ones derived by Braginskii in 1965 \cite{Braginskii_1965,FundamenskiEA07,GarciaEA05}. This fluid approach allows a simplified description by using macroscopic quantities that globally describe the system. 

The equations that govern our model are based on the drift-reduced Braginskii's equations. These represent the time evolution of the plasma density $n$, temperature $T$, and vorticity $\Omega$, in a simplified, 2D geometry (in a plane perpendicular to the main (toroidal) magnetic field.) The system is described by the following set of equations
\begin{gather}
    \label{continuity_eq}
	\frac{dn}{dt} + n C(\phi) - C(nT) = \nu_n \nabla^2 n - \sigma_n(x)(n-1), \\
	\label{energy_eq}
	\frac{dT}{dt} - \frac{7T}{3} C(T) - \frac{2T^2}{3n} C(n) + \frac{2T}{3} C(\phi) = \notag \\
 	\nu_T \nabla^2 T - \sigma_T(x)(T-1), \\
	\label{vorticity_time}
	\frac{d \Omega}{dt} - C(nT) = \nu_\Omega \nabla^2 \Omega - \sigma_{\Omega}(x) \Omega, \\
	\label{vorticity_eq}
	\Omega = \nabla^2 \phi.
\end{gather}
In the above set the total time-derivative is $\frac{d}{dt}=\frac{\partial }{\partial t} + \mathbf{u_E} \cdot \mathbf{\nabla}$, a sum of the partial time derivative and the advection term due to the electric drift velocity, while $C(\bullet)$ indicates an operator due to the magnetic curvature, as described below. 
The model is based on local slab coordinates: Cartesian axes $x, y, z$ play the role of radial, poloidal and toroidal coordinates of the device, respectively. 

The terms of the type ``$\nu_f \nabla^2$'' represent diffusive terms, where $\nu_f$ are imposed dissipation coefficients. Such viscous terms are simply introduced for numerical stability reasons since they dissipate high-wavenumber activity. Analogously,  $\sigma_f(x)$ represents linear damping terms due to particle transport along open magnetic field lines in the SOL. These coefficients are assumed to be the same for all variables, except for the temperature field which is damped five times stronger due to the predominant loss of hot electrons in the SOL region. 

The damping term is empirically chosen to be
\begin{equation} 
  \label{sigma}
  \sigma(x) = \lambda  \left[1 + \tanh \left( \frac{x-x_{LCMS}}{\delta} \right ) \right ],
\end{equation}
here $\lambda$ is the amplitude of the linear damping coefficient, $\delta$ is a measure of the transition, and $x_{LCMS}$ is the position of the last closed magnetic surface.  This term allows for the dissipation of the particle flow that reaches the outer boundary of the simulation domain. See \cite{GarciaEA04} for more details on the model.

We assumed a magnetic field along the $\hat{\bf{z}}$ direction as
\begin{equation}
   \label{Magnetic_B}
   \mathbf{B} = B(x) \mathbf{b} = \frac{1}{1 + \epsilon + \zeta x} \hat{\bf{z}}, 
\end{equation}
where the radial gradient depends on the parameter $\zeta$. 
Using the expression of the magnetic field, the curvature operator used in Eq.s~(\ref{continuity_eq})-(\ref{vorticity_eq}) is given by
\begin{equation}
    C = -\zeta \frac{\partial}{\partial y}.
\end{equation}
Note, moreover, that the assumption of considering $\mathbf{u}_E$ as the dominant contribution to the plasma motion (\textit{drift approximation}) allows to consider the vorticity $\Omega$ as the curl of the electric drift in the plane perpendicular to $\mathbf{B}$ \cite{MilitelloEA12}, with $\Omega = \mathbf{b} \cdot (\nabla \times {\mathbf{u_E}}) = \frac{\nabla^2{\phi}}{B}$. Here $\mathbf{b}$ is the magnetic unit vector in Eq.~(\ref{Magnetic_B}). The equation for the vorticity (\ref{vorticity_eq}) has been therefore obtained considering the difference between the continuity equation for ions and electrons and making use of the above assumptions.

\begin{figure}[h!]
    \centering
    \includegraphics[width=1.0\linewidth]{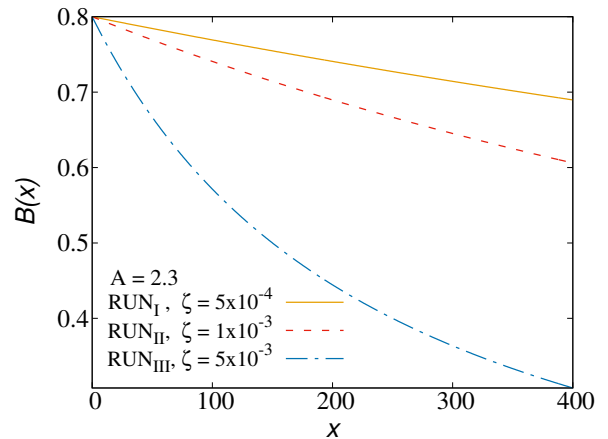}
    \caption{Magnetic fields profile in RUN-I, RUN-II, RUN-III. As the zeta parameter increases, the magnetic field drops faster along the radial coordinate. Hence the gradient of $\mathbf{B}$ becomes larger.}
    \label{B}
\end{figure}

The reduced Braginskii's equations are usually normalized considering the Bohm normalization where:
\begin{equation}
 \omega_{ci}t \rightarrow t, \quad \frac{\mathbf{x}}{\rho_s} \rightarrow \mathbf{x}, \quad  \frac{q \phi}{\mathcal{T}} \rightarrow \phi, \quad \frac{n}{\mathcal{N}} \rightarrow n,  \quad  \frac{T}{\mathcal{T}} \rightarrow T.
\end{equation}
Here $\omega_{ci} = q\mathcal{B}/m_i$ is the ion gyration frequency at a characteristic magnetic field strength $\mathcal{B}$, $\rho_s = c_s/\omega_{ci}$ the hybrid thermal gyration radius, $c_s = (\mathcal{T}/m_i)^{1/2}$ is the ion sound speed  and $\mathcal{N}$ and $\mathcal{T}$ are characteristic values of the particle density and electron temperature, respectively.  Taking $\mathcal{B}$ as  the  normalization of the magnetic field, the exact inverse toroidal field $\mathcal{B}/B =1+r cos \vartheta/ R $ is approximated by $1 + \epsilon + \zeta x $ in our non-dimensional units, where $\epsilon =r_0/R $ is the inverse aspect ratio. With this normalization, note that the gradient parameter $\zeta=\rho_s/R$, where $R$ is the major radius of the toroidal device.

\section{The numerical code}
\label{sec:numerical}
We developed a numerical algorithm that solves Eq.s~(\ref{continuity_eq})-(\ref{vorticity_eq}) by using a combination of finite differences and Runge-Kutta techniques for the integration. The 2D numerical code solves the equation in the ($x$, $y$) plane assuming periodicity in the $y$ direction. We used classical, second-order, centered finite differences for the spatial differentiation and second-order Runge-Kutta for the temporal evolution of the system. The code adopts a combination of finite differences and spectral methods to solve the Poisson equation, as commonly done in systems with a periodic direction.

\begin{figure}[h!]
    \centering
    \includegraphics[width=1.0\linewidth]{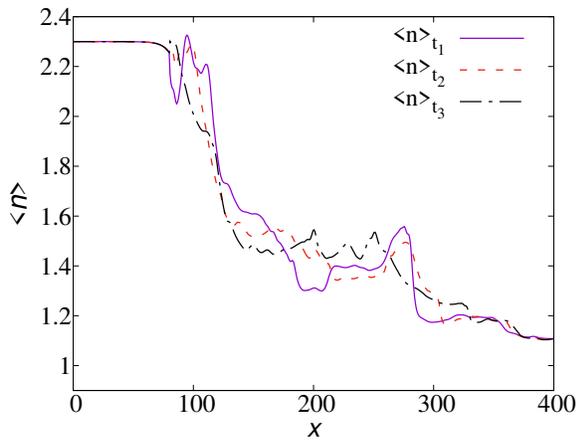}
    \caption{Picture of the average density in three subsequently moments showing that the initial mean profile is fixed over time in the inner region of the simulation domain (RUN-II).}
    \label{freezing}
\end{figure}
We applied the following boundary conditions in the radial direction $x$. At $x = 0$ (the inner boundary) $\phi = \Omega = 0$, and for the gradients $\frac{\partial n}{\partial x} =  \frac{\partial T}{\partial x} =  \frac{\partial \Omega}{\partial x} = \frac{\partial \phi}{\partial x} = 0$. For the outer boundary, at $x=L_x$, we chose $\phi = \Omega = 0$, $n=T=1$ and again null gradients.  Moreover, we have imposed on the potential that its poloidal average at $L_x$ is null.

As anticipated before, the domain consists of a rectangular box of size  $L_x \times L_y$ and a cell-centered grid with $N_x$ equidistant points in $x$-direction and $N_y$ equidistant points in $y$-direction.  In particular we used a box of $N_x \times N_y = 512 \times 256$ mesh points. We normalized these directions to the hybrid gyration radius $\rho_s=c_s/\omega_{ci}$, and in this normalized units $L_x = 400$ and $L_y = 200$. The radial position corresponding to the LCMS is located at $x_0 = 100$.

Regarding the shape of the magnetic field, it was important to set the parameters $\epsilon$ and $\zeta$. Following the literature, their values are usually small, and we set $\epsilon = 0.25$. The collisional coefficients $\nu_n = \nu_T = \nu_\Omega$ are set to a very low number ($\sim 10^{-2}$), in order to well-resolve the structures and dissipate artificial small scale fluctuations.  The linear damping term is assumed to be the same for the density and the vorticity but not for the temperature field, which is damped five times more, due to the predominant loss of hot electrons in the region of open magnetic field lines. In the expression of the linear damping, Eq.~\ref{sigma}, two other parameters have to be defined. In particular we used $\lambda = 1.6 \cdot 10^{-4}$ and $\delta = 1$. The parameters $\zeta$ and $\nu$ are varied for each Run, as described in the next Section.

In our numerical procedure, the transport of particles is ensured by maintaining constant in time the density and temperature mean profiles.  In this regard, we developed a new driving technique based on an averaging method: we forced the initial poloidal average of the density and temperature fields to remain constant within the LCMS region. The poloidal average of any field, is defined as the spatial average over the periodic direction, namely $\langle f \rangle(x) = \frac{1}{L_y}\int f(x,y') dy'$. In our model, at every time step, we subtracted the poloidal average of the particle density and temperature, replacing it with the initial profile. This procedure was applied only in the first part of the domain, up to the position preceding the radial coordinate of the LCMS, $x_0$. The method is very efficient and allows us to achieve a stationary configuration of emission/absorption. 

We named this driving technique, based on the freezing of the initial poloidal mean, \textit{average freezing driving} (AFD). This technique has been found to give very stably, statistically steady-state simulations of SOL turbulence \cite{ServidioEA08}.

\section{Numerical results}
\label{sec:results}

\begin{table}[ht]
	\begin{center}
		\makebox[0pt][c]{
		\begin{tabular}{ccccc}
		    \hline
			RUN & $\zeta$ & $A$ & $\Delta A$  & $\nu$ \\
			\hline
			$I$	& $5 \times 10^{-4}$ & 2.3 & 0.65   & $2 \times 10^{-2}$    \\ 
			$II$	& $1 \times 10^{-3}$ & 2.3 & 0.65  & $2 \times 10^{-2}$    \\
			$III$	& $5 \times 10^{-3}$ & 2.3 & 0.65   & $4 \times 10^{-2}$   \\
			$IV$	& $1 \times 10^{-3}$ & 1.8 &  0.40   & $2 \times 10^{-2}$  	 \\
			$V$	& $1 \times 10^{-3}$ & 3.3 & 1.15   & $5 \times 10^{-2}$ 	\\ 
			\hline
			    \label{tab1}
		\end{tabular}
	}
	\end{center} 
	\caption{Parameters used in the different runs. ``RUN'' indicates the simulation label, $\zeta$ the magnetic shear parameter, $A$ the edge fixed density, $\Delta A$ the relative density jump, $\nu$ the viscous coefficients. We used a timestep sufficiently small, with $\Delta t = 0.05$, writing every $100$ unit times. For the temperature driving, we set the same jump conditions as for the density ($A\equiv n_0=T_0$). The particle injection time $T_{0p}$ is set at 2500, while the total number of particles is $N=10000$.}
	\label{RUNS}
\end{table}

Thanks to the numerical code discussed in the previous Section it is possible to mimic the turbulent dynamics that characterize the outer region of fusion devices.  Here we report the main results of our numerical simulations.

We performed a campaign of runs, here summarized in TABLE \ref{RUNS}, where we vary mainly two important parameters, namely (1) the inner boundary values of the density and the temperature, $n$ and $T$,  and (2) the magnetic shear parameter $\zeta$. These changes allow us to inspect different turbulent regimes and to study the variations in the blobs' dynamics. For the main simulation, RUN$_{II}$, we have chosen an amplitude $A\equiv n_0=T_0$ of 2.3 and $\zeta = 10^{-3}$. The latter values are in agreement with previous works of the reduced Brangiskii model.

We analyze the outcome of the simulations via classical statistical analysis of turbulence, using both the Eulerian and Lagrangian approaches, as described in the next subsections.

\subsection{The Eulerian approach}
In order to characterize the dynamics of SOL turbulent plasmas, is important to define some relevant global quantities. The radial profile of any field, denoted hereafter by a ``zero'' subscript, is defined as its spatial average over the periodic direction $y$ (as done for the driving technique.) We can define the density and the poloidal flux profiles as \cite{Garcia_2001}
\begin{gather}
      n_0(x,t) = \frac{1}{L_y} \int^{L_y}_0 n(x,y,t) \; dy,  \\
     \nu_0(x,t) = \frac{1}{L_y} \int^{L_y}_0 \frac{\partial \phi}{\partial x} \; dy.  
\end{gather}
The deviation from the mean profile of any field is identified as the spatial fluctuation and it is indicated with a tilde, i. e. $\tilde{n}(x,t) = n(x,t) -n_0(x,t)$. Note that, because of the driving at $x<x_{LMFS}$, the profile is fixed in time for the density and the temperature, namely $n_0=n_0(x, t=0)$ and $T_0=T_0(x, t=0)$.

\begin{figure}
    \centering
    \includegraphics[width=1.0\linewidth]{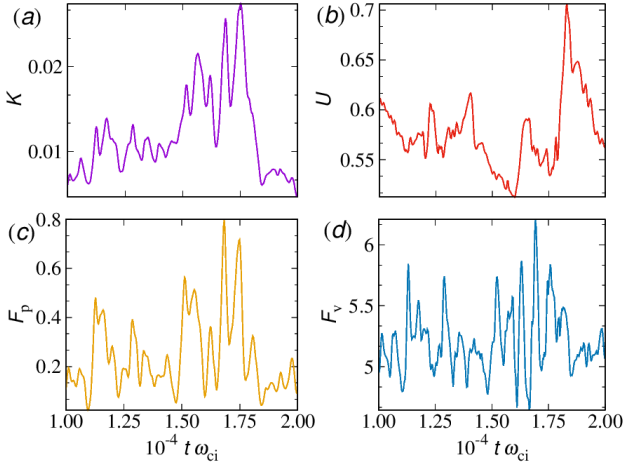}
    \caption{Kinetic $K$ and potential $U$ energies, energy transfer rates  ($F_p$ and $F_v$) during a turbulent burst event.}
    \label{burst_event}
\end{figure}

To characterize the turbulent state of the system in the outer layer, we study the evolution in time of the kinetic energy, considering its two components,  
\begin{equation}
\label{4.6}
 K = \int \frac{1}{2}(\nabla \tilde{\phi} )^2 d\mathbf{x}, \quad \quad U = \int \frac{1}{2}\nu_0^2 d\mathbf{x}.
\end{equation}
These are the kinetic energy in the fluctuating motions and the sheared poloidal flows, respectively. Following \cite{GarciaEA2003}, the evolution of these integrals is given by some straightforward manipulations of Eq.s ~(\ref{vorticity_time}) and (\ref{4.6}), leading to
\begin{gather}
  \label{kinetic}
  \frac{dK}{dt} = \zeta \int nT\tilde{u_x} \; d\mathbf{x} - \int \nu_0 \frac{\partial}{\partial x}( \tilde{u_x} \tilde{u_y})_0 \; d\mathbf{x} - \int \tilde{\phi} \Lambda_\Omega \;d\mathbf{x}, \\
  \label{poloidal}
  \frac{dU}{dt} = \int \nu_0 \frac{\partial}{\partial x}( \tilde{u_x} \tilde{u_y})_0 \; d\mathbf{x} - \int \phi_0 \Lambda_\Omega \; d\mathbf{x}.
\end{gather}
The first term of the Eq.~(\ref{kinetic}) corresponds to the fluctuating motion change due to the radially advective transport of thermal energy. The second one, together with the first term of Eq.~(\ref{poloidal}), represents the conservative transfer of kinetic energy from fluctuating motion to the sheared flow (i. e. kinetic energy may be transferred between the convective cells and sheared poloidal flows by a radial inhomogeneity in the off-diagonal components of the averaged Reynolds stress tensor). The last term present in both equations is linear damping due to collisions and particle transport.

The energy transfer rates from thermal energy to the fluctuating motion, and from fluctuating to the sheared flows, can be defined, respectively, by
\begin{equation}
\label{eq:fluxes}
    F_p = \zeta \int n T \tilde{u_x} \; d\mathbf{x}, \quad \quad F_v = \int \frac{\partial \nu_0}{\partial x}( \tilde{u_x} \tilde{u_y})_0 \; d\mathbf{x}.
\end{equation}
These fluxes are crucial for the characterization of turbulent transport, as we shall see in a little.

\begin{figure}
    \centering
    \includegraphics[width=1.0\linewidth]{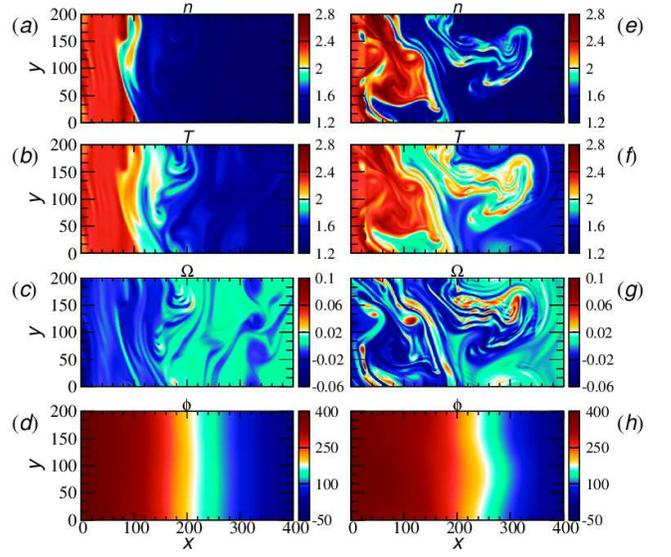}
    \caption{Comparison of the spatial structure of particle density, temperature, vorticity, and potential during a quiet period, pictures(a)-(d), and during a burst event, (e)-(h).}
    \label{quiet_burst}
\end{figure}

The energies in Eq.~(\ref{4.6}) and the fluxes in Eq.~(\ref{eq:fluxes}) are highly representative of the system dynamics. Hereafter we show the results for the main RUN$_{II}$. The relation between all the above global quantities is better represented in figure \ref{burst_event}, where concentrate on a single ``burst event''. In correspondence with blob ejection, there is a rapid growth of the energy transfer rate $F_p$ since the thermal energy exchange its energy with the fluctuating motion $K$. Then, after a short time, there is a rapid decrease of the latter in favor of the sheared poloidal flow. A peak in the quantity $F_v$ at the same subsequent time is observed. Both the trend of the kinetic energy and that of the energy transfer rate suggest that the SOL is characterized by the alternation of blob ejection moments and quiet periods.

The intensity of the fluctuating motions shows irregular oscillations with pronounced bursts throughout the simulation. This is a typical manifestation of intermittent, bursty turbulence, typically observed in many tokamak devices. During such burst events, the formation of blobs takes place. Whenever this fluctuation level is sufficiently large, there is a rapid growth of the energy in the sheared poloidal flows, followed by a suppressed level of fluctuation kinetic energy. Such a dynamic is the consequence of the exchange of kinetic energy from the fluctuating motion to the sheared flows.

The difference between an emission moment and a quiet period is reported in figure \ref{quiet_burst}. In the quiet regime, \ref{quiet_burst}$(a)$, the SOL region is characterized by almost no spatial fluctuations and the plasma is nearly homogeneous and isothermal. There are plumes in the particle density and temperature fields. At the time of a burst event [(e)-(h)] it is evident that prominent structures develop in the SOL. During such a burst event, an accumulation of particle density and heat is clearly evident. This ejection is localized in the region of the last closed magnetic surface, in the SOL, at $x>x_{LMCS}$. 

\begin{figure}
    \centering
    \includegraphics[width=1.0\linewidth]{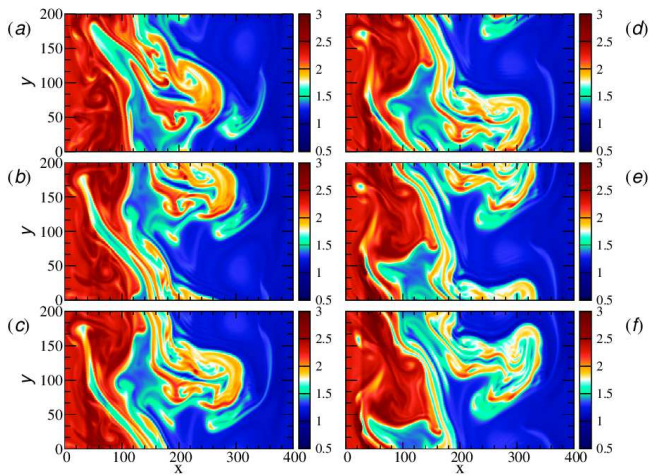}
    \caption{Spatial evolution of particle density over time, from (a) to (f), in sequence, showing a blob growth from the edge of the plasma column and its propagation towards the outer boundary. }
   \label{blob_evolution}
\end{figure}

During a burst event, it is possible to observe the formation of a blob by following the evolution over time of particle density, as can be seen from figure \ref{blob_evolution}. These plots show the rapid development of a structure from the inner region to the outer part of the domain, i. e. towards the open wall. During the emission, the main vortex can also merge and eat small neighboring fluctuations, as can be observed from the time history of the emission process.

\begin{figure}
    \centering
    \includegraphics[width=0.9\linewidth]{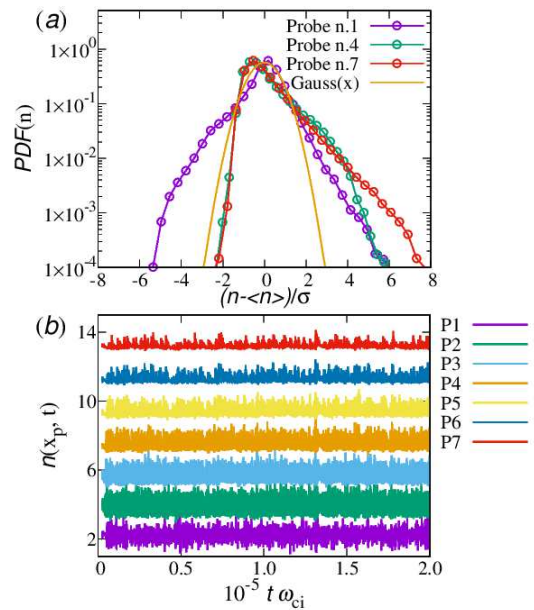}
    \caption{(a) PDF of the particle density signals, compared for three different probes; (b) Temporal evolution of the particle density measured by the different probes. Note that the signal has been shifted in the y-axis in order to compare on the same plot all the measurements.}
	\label{PROBES-PDF}
\end{figure}

Another useful Eulerian analysis that can help to understand the propagation in space and time of turbulent structures is the analysis of macroscopic variables such as the density at selected regions of the numerical domain (to mimic probe measurements in the SOL). We measured the density and the probability distribution function (PDF) of the particle density at different points (probes) of the simulation domain. These probes are 7 points equally spaced, numbered going from left to right of the domain (so the first probe is situated in the plasma edge region, while the last one is near the wall).

In figure \ref{PROBES-PDF}-(a), we report the PDF of the particle density, at three different probes ($P_1$, $P_4$ and $P_7$). The presence of fluctuations causes a departure of the PDF from the normal distribution, manifesting very high exponential tails \cite{GarciaEA2003, AntarEA2001, SattinEA05}. This is typical of intermittent, extraordinary events. As we can observe in figure \ref{PROBES-PDF} $(a)$, at the first probe which is located within the plasma edge, the $PDF(n)$ has a nearly-symmetric, super-Gaussian shape. At larger distances from the LCMS, due to the intermittent nature of turbulence, it gradually develops higher tails, skewed in the outward direction. This skewness and kurtosis of the distribution is commonly observed in SOL of fusion devices and reflects the high probability of extreme events, as largely investigated in the literature \cite{AntarEA03, XuEA05, vanMilligenEA05, ArakawaEA2010}.

The panel $(b)$ highlights the change in time of the density at different points of the domain. The signal is stronger in the first probes since the greatest concentration of density occurs at the inner edge and decreases near the wall, at probe $P_7$. The figure shows small signal peaks that interchange between one probe and the other, becoming less and less intense. We can see that some events are correlated. These effects are due to the turbulent transport and reveal that blobs structures are moving radially towards the wall. The particle density at the probes within the edge region, $P_1$, and $P_2$, shows chaotic oscillations which appear symmetric concerning the mean value, while all other probe signals reveal predominantly positive fluctuations, as highlighted from the PDF analysis,  characterized by fast rise followed by damped irregular oscillations. 

\begin{figure}
    \centering
    \includegraphics[width=1.0\linewidth]{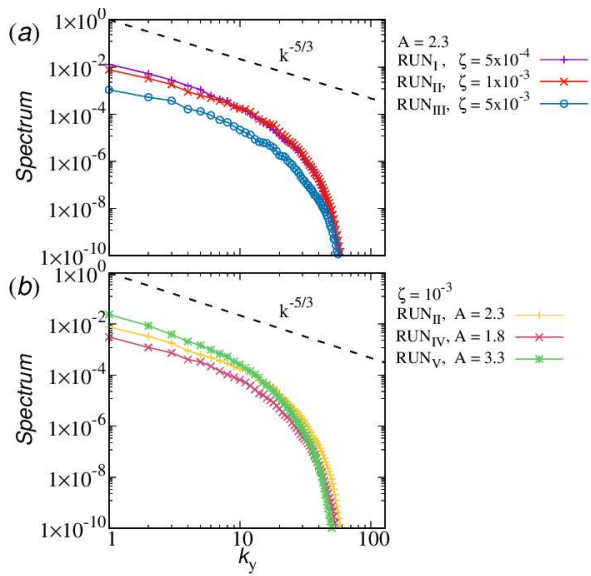}
    \caption{(a) Comparison between the average energy spectrum of RUN-I, RUN-II, and RUN-III. The energy spectrum decreases as the $\zeta$ value increases, revealing a more stable configuration in the RUN-III, which is characterized by a higher value of $\zeta$; (b) Comparison between the average energy spectrum of RUN-II, RUN-IV, and RUN-V. In this case, the energy spectrum decreases in the presence of a smaller pressure gradient, reflecting the hypothesis that a higher gradient of density injects more stationary turbulence into the system.}
	\label{spectrum}
\end{figure}

\begin{figure}
    \centering
    \includegraphics[width=1.0\linewidth]{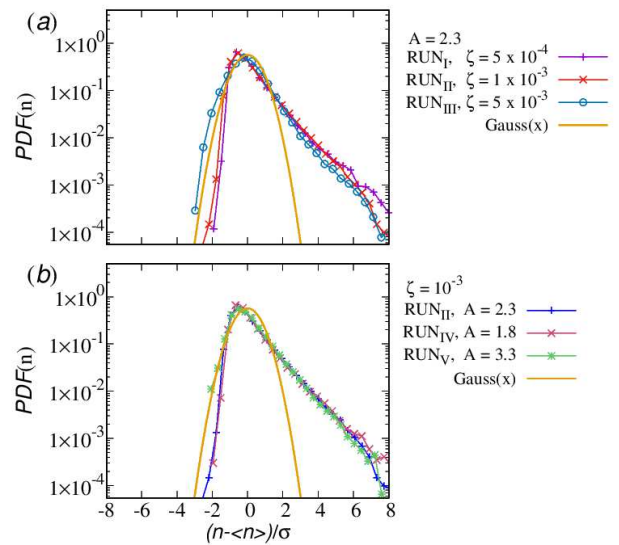}
    \caption{(a) Comparison of the probability distribution function of particle density between RUN-I, RUN-II, and RUN-III. More pronounced tails are present in the RUN-I and RUN-II rather than in the RUN-III, indicating more intermittency in the former cases; (b) Comparison of the probability distribution function of particle density between RUN-II, RUN-IV, and RUN-V. The patterns are quite the same revealing that density and temperature gradients do not influence too much the nature of the intermittent transport.}
	\label{PDFs}
\end{figure}

At this point, we change the simulation settings to explore the parameter space. By changing the $\zeta$ and $A$, we analyze different turbulent regimes and we compare all the simulations by evaluating their power spectra and the PDFs of the particle density. For the power spectra, we Fourier-transform the density field $n(x,y,t)$ along the periodic direction, compute the energy spectrum $E(x, k_y, t)=|\tilde{n}(x, k_y, t)|^2$ ($\tilde n$ being the Fourier coefficient), and finally obtain the spectrum from an  ensemble-average both in time and space (in the region $x>x_{LMCS}$).

In RUN$_I$, RUN$_{II}$, and RUN$_{III}$ we have increased progressively the value of $\zeta$. Since this parameter governs the variation along the radial coordinate of the magnetic field, we have effectively increased the gradient of $\mathbf{B}$ as can be evinced from figure \ref{B}.

In figure \ref{spectrum}-(a) we compare the averaged spectra of the plasma density $n$, as a function of the poloidal mode $k_y$, for this campaign of runs. The turbulent spectrum progressively decreases from RUN$_I$ to RUN$_{III}$. The spectra all manifest a power law scaling, consistent with a fluid-like turbulent cascade \cite{KOLMOGOROV_41a, KOLMOGOROV_41b}. We report also the Kolmogorov scaling law as a reference. Even if the scaling is consistent from one run to another, the amplitude of turbulence decreases.  This trend reveals that an increase in the value of $\zeta$ leads to more stable configurations, while the system characterized by a small value of $\zeta$ is affected by more turbulent transport. The suppression of turbulent transport is also evident in the PDFs in figure \ref{PDFs}-(a). The low-$\zeta$ cases are more turbulent and intermittent. 

RUN$_{II}$, RUN$_{IV}$, and RUN$_{V}$ are characterized by different values of the initial density and temperature $A$. Changing the value of these quantities at the inner boundary of the domain means changing the gradient of pressure that, in consequence of the inhomogeneity of the magnetic field $\mathbf{B}$, drives instabilities towards the outer boundary region. From the spectra in figure \ref{spectrum}-(b) it is clear that higher density and temperature gradients (higher $A$) cause more turbulent activity. The intermittent behavior seems not affected so much by the change of $A$, as it can be seen from the  PDFs in figure \ref{PDFs}-(b).

This picture confirms that the available source of free energy, driving the convective motions, is contained in the mean pressure gradient. An increase in the free energy leads to an increase in the fluctuating motion and, consequently, to a higher turbulent (radial) transport.

\subsection{The Lagrangian approach}
\begin{figure}[h!]
    \centering
    \includegraphics[width=1.0\linewidth]{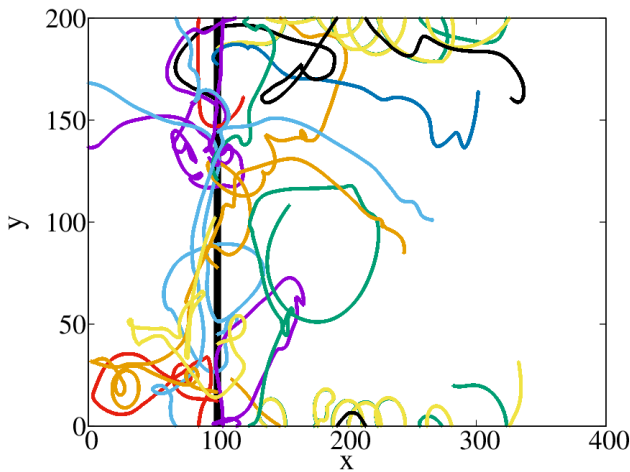}
    \caption{Examples of trajectories of the fluid elements, for RUN-II, starting from the initial position at $x_0$ (here highlighted by a black vertical line). Only early times of such trajectories, for a limited sample, are reported.}
    \label{trajectories}
\end{figure}
In the second part of our analysis, to better understand the diffusion of fluid elements in the SOL region and characterize the transport dynamics, a new technique is investigated, based on the Lagrangian approach. 

We followed in time fluid tracers in the drift approximation, integrating the trasjectory of the plasma element in the numerical domain. In practice, we solve particle integration in our code, following the trip of plasma elements from the central part of the LCMS towards the colder, empty regions at the wall. We solved the fluid equation of motion in which we consider only the major contributions of the cross-field fluid drift -- the electric and the diamagnetic drifts.

\begin{figure}[h!]
    \centering
    \includegraphics[width=1.\linewidth]{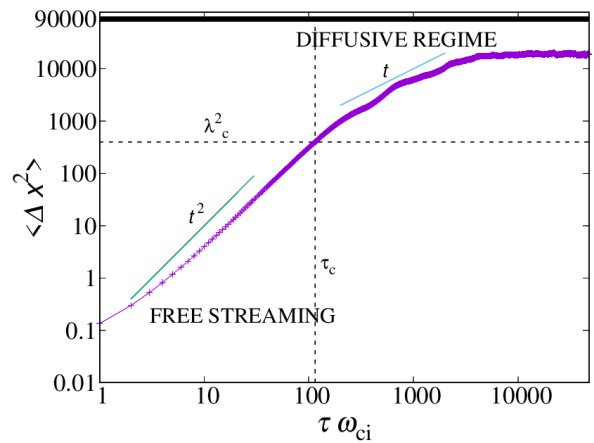}
    \caption{Temporal evolution of the mean square displacement of the fluid elements, in a double logarithmic scale, for RUN-II. The horizontal (dashed) line represents the square of the correlation length $\lambda_c$, namely the average size of the blobs. The characteristic decorrelation time is reported with a vertical dashed line. The thick (black) line represents the maximum allowed MSD, $\Delta x_{max}^2$.}
    \label{particles}
\end{figure}

The equation of motion for the fluid tracers has been derived by using the same Braginskii approximation of our model: we neglect ions dynamics, concentrating only on the motion of the electrons, and we omit the inertial term in the cross-field drift. The drift velocity then is simply
\begin{equation}
    \label{eq:u}
    \mathbf{u_\perp} = \mathbf{u_E} + \mathbf{u_d} = \frac{1}{B}  \mathbf{b} \times \nabla \phi - \frac{1}{qnB} \mathbf{b} \times \nabla p.
\end{equation}
Again, we normalized this expression following Bohm's normalization introduced previously. Plasma fluid elements will therefore move with the above velocity, at each time, by obeying $\mathbf{\dot{x}} = \mathbf{u_\perp}$. We integrated the trajectories of fluid particles that follow passively the flow. In this regard, we upgraded the numerical code with this Lagrangian tracing technique where the fields have been continuously supplied by the time-varying simulation, and by obtaining the velocity (\ref{eq:u}) at each particle position via a  bilinear interpolation algorithm (second order in space).

At the beginning (or at the earlier stages) of each simulation, we placed all the Lagrangian tracers at the LCMS $x_0$, but with different (random) poloidal positions $y$. We considered the following boundary condition for the tracers: if an element reaches the radial boundary, we remove it from the domain since we are interested in particles within the SOL region and, in a fusion device, a particle that interacts with the wall is lost through the divertor. In the poloidal direction, instead, we consider periodic boundaries and a particle that reaches the bottom edge, for example, reappears from above. The parameters of the Lagrangian integration are summarized in Table I and their initial motion is represented in figure \ref{trajectories}. We show the evolution in time of the trajectories for just 10 fluid elements, to avoid overcrowding. In this figure, one can see how, from the radial position corresponding to the LCMS, elements start their erratic trip in the simulation domain under the effect of turbulence. Some of them are absorbed in the left part of the wall, namely inside the device. Part of it is lost in the outer SOL (right side).

\begin{figure}[h!]
    \centering
    \includegraphics[width=1.\linewidth]{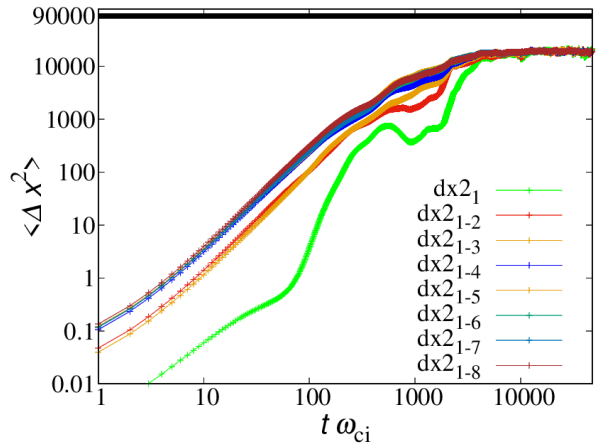}
    \caption{Convergence of the statistical analysis, for RUN$_{II}$. Other runs have similar behaviors. }
    \label{dx2}
\end{figure}

To make a statistical analysis of the turbulent diffusion mechanism, we consider a large sample of fluid elements, $N = 10000$, and calculate their displacement from the initial position $\Delta x(t)= x(t)-x_0$, along the $x$ axis. Note that with this setup, the maximum value of the displacement is $\Delta x_{max} = 300$. We verified the convergence of the statistics presented below by increasing systematically the number of particles, as reported later in this Section. We noticed that plasma elements in the SOL region experience an irregular random walk.

To make contact with classical diffusion theories, we evaluate the mean square displacement (MSD) of the radial positions $\langle \Delta x \rangle(t)$, where $\langle \bullet \rangle$  represents an ensemble over all the particles. In figure \ref{particles}, we report the MSD as a function of time.  The behavior of the MSD is quite interesting, given the fact that we are in a small, finite-size system. In the initial stage of their journey, the particles experience a behavior consistent with $ \tau^2$, typical of free-streaming motion \cite{HuangEA11, Pusey_2011, Taylor_1920}. Here  $\tau$ is simply the difference between $t$ and the injection time $T_{0p}$ (see Table I). After the above initial transient, $\langle \Delta x^2\rangle \sim \tau$, the curve suggests a diffusive behavior. As in typical plasma turbulence works \cite{PecoraEA18}, this regime starts when the displacement becomes on the order of the correlation length of turbulence $\lambda_c$. The latter corresponds to the typical size of the largest energy-containing eddies, namely the size of the blobs. We estimated this length to be qualitative $\lambda_c \sim 20$, for the above simulations. The regime of the diffusive transient starts therefore at a characteristic correlation time $\tau_c$, highlighted in the figure with a vertical line. At later times, contrary to unbounded problems of diffusion, the behavior is influenced by the system size, and the large-scale diffusive behavior is lost. At these large times, most of the particles abandoned the domain, either because are absorbed by the inner boundary, or because they are lost at the outer wall.

\begin{figure}[h!]
    \centering
    \includegraphics[width=1.\linewidth]{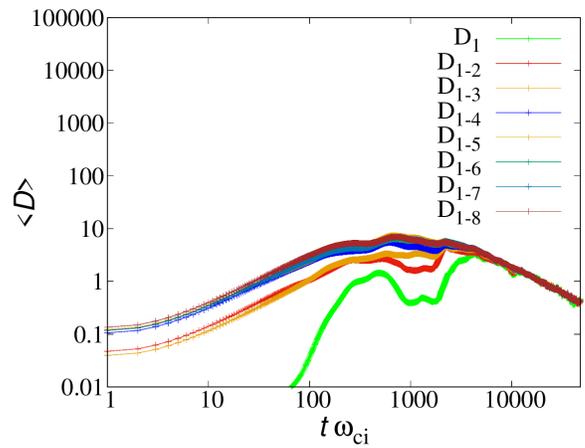}
    \caption{Temporal convergence of the diffusion coefficient, estimated as $ D = \langle \Delta x^2\rangle /(2 t)$ (RUN-II). }
    \label{diffusion-D}
\end{figure}

To further confirm this diffusive behavior, we computed the diffusion coefficient, but first we checked the convergence of our results by increasing systematically the statistics. We divided the simulation times into subdomains. In particular, we divided the particle integration into eight, smaller, subsequent shots, each of which lasts $5\times10^{4}$ computational times. At every particle restart, we set all the fluid elements at the initial position ($x_0 = 100$) and we studied how the diffusion changes over time.  In figure \ref{dx2} we show the temporal convergence of the MSD: ``$dx2_1$'' refers to the MSD of the first part of the simulation, the second, $dx2_{1-2}$, refers to the average of the first and second part of RUN-II, and so on. We continue to enlarge the statistics until we comprehend all the eight time subdomains. We can observe that the curves progressively increase, until they collapse, suggesting a clear convergence of the statistics.

At this point, the diffusion coefficient can be computed as $ D \sim \langle \Delta x^2\rangle /(2 t)$, for each average shown in figure \ref{dx2}. In particular, as it can be seen from figure \ref{diffusion-D}, the diffusion coefficient saturates to a single curve when all the time-shots of the particles are taken into account. Note that this happens already when just 6 datasets are averaged, confirming that the total acquisition time of the experiment is statistically relevant.

\begin{figure}[h!]
    \centering
    \includegraphics[width=1.\linewidth]{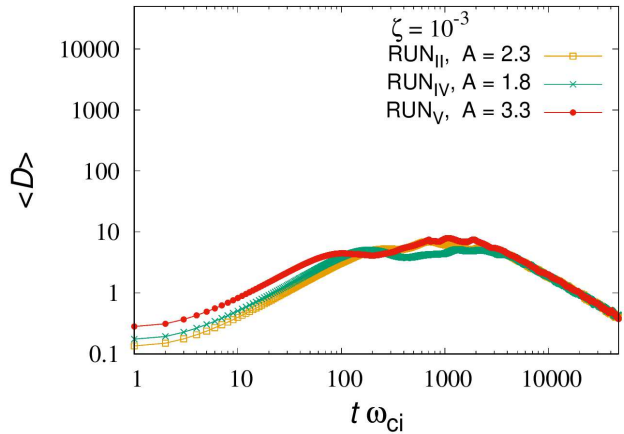}
    \caption{Comparison of the temporal evolution of the diffusion coefficient in a double logarithmic scale between RUN-II, RUN-IV, RUN-V }
    \label{diffusion-Dcompare}
\end{figure}

Finally, we compare the diffusion coefficient, averaged over all the datasets, for the runs where we have a strong enhancement of turbulence, namely by comparing RUN$_{II}$, RUN$_{IV}$, and RUN$_{V}$. In figure \ref{diffusion-Dcompare} one can see that diffusion is effectively higher for the run with stronger turbulence, namely the case where the density and temperature driving is more intense ($A=3.3$). This preliminary analysis of the diffusion process further confirms that turbulence enhances particle diffusion, even in such a simplified Brangiskii model of the SOL.

\section{Conclusions}
\label{sec:conclusions}
A two-dimensional, simplified model has been adopted to study the turbulent dynamics in the region of the scrape-off layer. We show how the numerical model, based on the reduced Braginskii equations, can describe the formation and the evolution over time of blobs structures. We study these complex dynamics by using both classical Eulerian analysis and the Lagrangian approach, by varying the ambient conditions of the plasma. 

We conducted a simulation campaign where we varied the magnetic shear intensity and the level of injected turbulence. We proposed a new technique for driving the bursty turbulence based on the eternal freezing of the main plasma profiles in the inner region of the SOL. As in previous works, we observed an intermittent character of turbulence, where the latter is composed of blobs of plasma that propagate outward, from the LCMS toward the walls. The radial flux is dominated by mushroom-like blobs that move with the $E\times B$ drift. The statistical analysis of the flux, as measured by single probes in the domain, is in agreement with experimental observations, manifesting highly non-Gaussian PDF, characterized by intermittent, extreme events.

By varying the simulation parameters, in a steady state regime of driven turbulence, we observed that the turbulence level is higher for smaller coefficients $\zeta$ of the Brangiskii model. Similarly, the turbulence intensity increases with larger density and temperature jumps, going from the internal to the outer region.

In the second part of the work, we inspected the diffusive properties of the plasma by integrating fluid trajectories, under the fluid-drift approximation. Following a large sample of elements, we observed a diffusive transient, for a length-scale larger than the typical blob size.

This work is relevant for the comprehension of the turbulent transport at the edges of tokamak devices. Future studies will be devoted to the modeling of larger domains and with higher resolutions. A more detailed theoretical investigation of the diffusive process will be presented in other works. The effect of neutrals on the dynamics of the system will be investigated as well in upcoming studies.

\end{document}